\documentclass[format=sigconf,letterpaper]{acmart}
\usepackage{multirow,multicol}
\usepackage{color,tabularx, colortbl,amsmath,bm,arydshln}
\usepackage{enumitem,kantlipsum,arydshln}
\usepackage{booktabs,url,tcolorbox,mathtools} \usepackage{newtxmath}

\acmConference{Submitted to CIKM 2022}{October 17-21, 2022}{Atlanta}
\acmBooktitle{Proceedings of the 31st ACM International Conference on Information and Knowledge Management}
\copyrightyear{2022}
\acmYear{2022}
\setcopyright{acmlicensed}
\acmPrice{XX.XX}
\acmDOI{XX.XXXX/XXXXXXX.XXXXXXX}
\acmISBN{XXX-X-XXXX-XXXX-X/XX/XX}

\begin{document}

\title{Unsupervised Question Clarity Prediction\\ Through Retrieved Item Coherency}

\author{ Negar Arabzadeh }
\email{narabzad@uwaterloo.ca}
\affiliation{%
   \institution{University of Waterloo}
   \country{Canada}
   }

\author{Mahsa Seifikar }
\email{mahsa.seifikar@uwaterloo.ca	}
\affiliation{%
   \institution{University of Waterloo}   \country{Canada}
}

\author{ Charles L. A. Clarke }
 \email{charles.clarke@uwaterloo.ca}
\affiliation{%
   \institution{University of Waterloo}   \country{Canada}
}

\begin{abstract}
Despite recent progress on conversational systems, they still do not perform smoothly and coherently when faced with ambiguous requests. When questions are unclear, conversational systems should have the ability to ask clarifying questions, rather than assuming a particular interpretation or simply responding that they do not understand. Previous studies have shown that users are more satisfied when asked a clarifying question, rather than receiving an unrelated response. While the research community has paid substantial attention to the problem of predicting query ambiguity in traditional search contexts, researchers have paid relatively little attention to predicting when this ambiguity is sufficient to warrant clarification in the context of conversational systems. In this paper, we propose an unsupervised method for predicting the need for clarification. This method is based on the measured coherency of results from an initial answer retrieval step, under the assumption that a less ambiguous query is more likely to retrieve more coherent results when compared to an ambiguous query. We build a  graph from retrieved items based on their context similarity, treating measures of graph connectivity as indicators of ambiguity. We evaluate our approach on two recently released open-domain conversational question answering datasets, ClariQ and AmbigNQ, comparing it with neural and non-neural baselines. Our unsupervised approach performs as well as supervised approaches while providing better generalization.

\end{abstract}


\maketitle

\section{Introduction}

\begin{figure}[t!]
\centering
\includegraphics[width=0.42\textwidth]{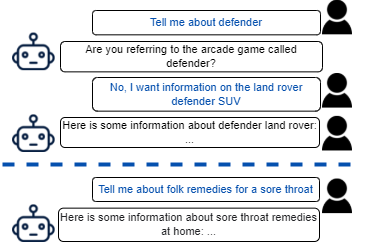}
\caption{Examples of two open-domain requests between a user and an intelligent system from ClariQ dataset \cite{aliannejadi2021building}. The top conversation requires a clarifying question due its ambiguty. In the bottom conversation, the user's query is not ambiguous and the retrieved items are provided to the user immediately .}
\label{fig:running_example}
\end{figure}
Recently, the Information Retrieval (IR) and Natural Language Processing (NLP) communities have been enjoying major performance enhancements in many core tasks including ad hoc retrieval, question answering, and conversational search~\cite{gao2022neural,lin2021pretrained,mitra2018introduction,karpukhin2020dense,craswell2021trec}.  Despite this recent progresses, even the best-performing systems are not able to perform smoothly and coherently on all requests \cite{lipani2021doing,mackie2021deep,arabzadeh2021ms}. For example, broad queries can have multiple interpretations and aspects~\cite{cronen2002quantifying,cronen2002predicting,song2007identifying,roul2012effective}, and satisfying such ambiguous requests has long been considered as a challenging task. Consequently, the research community has paid substantial attention to the problem of predicting query ambiguity in traditional search contexts~\cite{clarke2009effectiveness,wang2010query,sanderson2008ambiguous,kadir2018semantic}.
In some contexts, such as web search, the impact of query ambiguity can be alleviated with approaches such as diversification or session-based data~\cite{drosou2010search,vallet2012personalized}. However, in some contexts, such as smart assistants, a single response is required, and it is not possible to adapt solutions such as diversification.

An alternative approach asks a clarifying question before responding~\cite{aliannejadi2019asking,rao2018learning,stoyanchev2014towards}.  Studies have shown that when a user has a broad or 
ambiguous request, they will be more satisfied if they are asked a clarifying question~\cite{zamani2020generating,kiesel2018toward}.  
\citet{zamani2020generating} conducted a user study on clarifying questions, showing that users appreciate clarifying questions because they are functional and give the user a sense of confidence.  People anticipate having conversations with intelligent systems that are similar to the conversations they might have with humans \cite{nass2000machines}. When humans are having a hard time understanding each other, they tend to ask questions to make their understanding clearer. 
Thus, asking clarifying questions is essential for developing human-like dialogue systems.
All in all, when questions are unclear, conversational systems should have the ability to ask clarifying questions, rather than assuming a particular interpretation or simply responding that they do not understand. 

A key challenge is determining when query ambiguity is sufficient to warrant clarification. Consider the two conversations in Figure \ref{fig:running_example}. In the first example, the user issues an ambiguous query (\textit{``Tell me about defender''}). The system identifies the request as ambiguous, and initiates a clarifying question to narrow down the user's intent. After the user clarifies their intent, the system is able to provide appropriate information. In the second example, the query (\textit{``Folk remedies for a sore throat''}) is sufficiently clear that no clarification is required. Since the system do not consider the query as the ambiguous one, it proceeds to retrieval step right-away without asking any clarifying question.
In this paper, our overall goal is to  differentiate between the two cases exemplified in Figure \ref{fig:running_example}, i.e., to determine when the system should ask a clarifying question. 

Researchers in the IR and NLP communities have extensively studied ``What to ask?''  when clarifying a user's intent. They have employed traditional approaches, as well as neural models, such as transformers, sequence-to-sequence models, and recurrent neural networks to generate clarifying questions~\cite{zamani2020generating,aliannejadi2021building,aliannejadi2019asking,braslavski2017you,rao2018learning,stoyanchev2014towards,de2003analysis,10.1007/978-3-030-99736-6_28,sekulic2021towards}. However, before we determine what clarifying question to ask, we must decide when a clarifying question is required. This problem of determining ``When to ask?'' has received relatively little attention.
While these works are much appreciated, the initial step for having a smooth conversation with intelligent systems is to determine the moment a system should ask clarifying question. 
To the best of our knowledge, unlike the substantial research that is conducted on "what to ask", "When to ask" clarifying questions has not been extensively studied yet. In this work, we focus on when the query is considered as an ambiguous one and clarification questions are required.
Addressing this problem requires significant user understanding, which makes test collection development for this task expensive and labor-intensive.  As a result, there are few datasets suitable for studying this problem.
This fact  explains why there are not many available datasets for studying clarifying questions. 
\citet{aliannejadi2021building} were among the pioneers attempting to formulate and curate data for when clarifying questions should be asked. They collected and released a comprehensive dataset,  ClariQ, with a focus on asking clarifying questions in open domain conversational systems.  ClariQ includes both ambiguous and unambiguous queries. In creating this dataset, they asked human annotators to assign a score for each request based on the level of clarification each query required. Current state-of-the-art query clarity prediction methods on the ClariQ dataset employ fine-tuned large pre-trained language models, such as BERT, to predict this clarification level~\cite{aliannejadi2021building}. However, due to limited available training data, this approach could be potentially biased toward that training set and lack generalizability. 

To avoid possible problems related to bias and lack of available training data, we propose an unsupervised method to address query clarity prediction based on the coherency of the retrieved items from the initial user request. Similar to previous query performance prediction methods \cite{carmel2010estimating,he2008using,zhao2008effective,arabzadeh2021query}, we base our work on the intuition that a more clear query is more likely to retrieve coherent results when compared to an ambiguous one. The more ambiguous the query is, the more varied the returned items will be. Thus, inspired by  coherency metrics \cite{putra2017evaluating,tsekouras2017graph}, we build a graph from retrieved items based on their coherency, treating measures of graph connectivity as indicators of ambiguity. We distinguish our work from other coherency measures in several ways:
1) We deploy coherency metrics in the new context of clarifying questions. 
2) We focus on the coherency of retrieved items at the passage level, using contextualized-based embeddings rather than sentence-level similarity 3) We measure coherency by treating passages as potential successors, i.e., we consider a pair of documents to be coherent if a pre-trained language model predicts that they could be successive passages.
We evaluate our proposed approach on the ClariQ open-Domain dialogue corpora, comparing it with current SOTA baselines~\cite{aliannejadi2021building}. In addition, we evaluate the performance of SOTA post-retrieval query performance predictors on predicting query clarity level~\cite{carmel2010estimating,tao2014query,zhou2007query,shtok2012predicting}.
Further, we investigate the generalizability of supervised SOTA methods and compare them with our proposed method by applying the query clarity measures in a zero-shot setting to the AmbigNQ dataset \cite{min2020ambigqa}. AmbigNQ is a recently released dataset with annotations on NQ-OPEN questions
containing diverse types of ambiguity \cite{kwiatkowski2019natural}.
We show that while our unsupervised approach outperforms supervised methods and exhibits greater generalizability.
\section{Related Work}

Clarifying questions have proven their value in a number of search settings. \citet{zamani2020generating} explored users' clarifying questions when searching on the web. These users described clarifying questions in search platforms as convenient, as they save time and steps. \citet{kiesel2018toward} studied the effect of voice query clarification and concluded that users like to be prompted for clarification.

"What to ask?" has seen substantial recent attention in the IR and NLP communities. For example, \citet{aliannejadi2019asking} formulated the task of asking clarifying questions in open-domain information-seeking conversational systems and compiled the Qulac dataset, which comprises more than 10K question-answer pairs for multi-faceted topics. They hired human annotators to create clarifying questions. While one way of asking clarifying questions is to select them  form a pool of generated questions,  \citet{zamani2020generating} took another approach by training a sequence-to-sequence model and recurrent neural networks on query reformulations as weakly supervised training data. Datasets such as MIMICS and QULAC are considered as functional outcomes of such work.
They include multi-faceted queries accompanied with user behavioural signals. Such datasets provide a suitable vehicle for other researchers to train  clarifying questions models~\cite{sekulic2021towards}. Further, \citet{10.1007/978-3-030-99736-6_28} addressed limitations of previous works by extracting useful facets and proposing a facet-based clarifying question generation approach. 

We determine when the system should ask clarifying questions with a graph-based approach.
Graphs have long demonstrated their potential for hosting textual similarity measures. \citet{DBLP:conf/ranlp/TsekourasVG17} present a graph-based text similarity that takes advantage of named entities, boosting the performance of text clustering tasks. Similarly, \citet{putra-tokunaga-2017-evaluating} propose graph construction methods based on the semantic similarity of vertices for document discrimination tasks based on text coherency. Since graph-based coherency metrics have shown to be effective in other tasks, we adapt graph-based coherency metrics to determine query clarity. 
\begin{figure*}[!h]
\centering
\includegraphics[width=0.92\textwidth , clip, trim=2cm 3.8cm 2cm 8.5cm,scale=0.5]{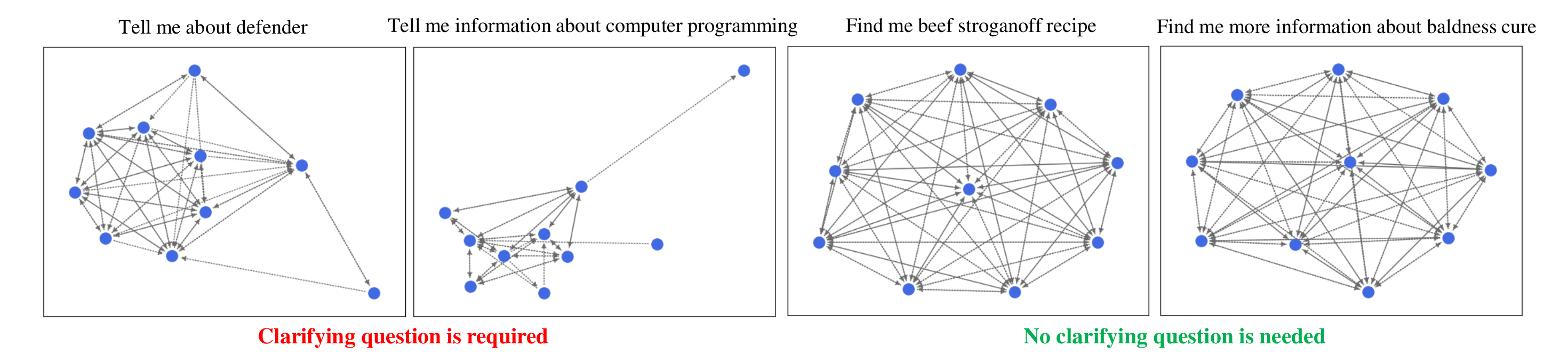}
\caption{Examples of coherency networks for ambiguous and non-ambiguous queries. }
\label{fig:graphs}
\end{figure*}

\section{Proposed Approach}
\label{proposed}
We base our approach on the intuition that an ambiguous query will retrieve items that refer to different aspects of the query. For example, consider the query \textit{"Tell me about defender"} (Figure \ref{fig:running_example}). Here, ``Defender'' could refer to the Land Rover Defender, the 1981 Defender arcade game, Microsoft Defender, or other things. Thus, retrieved items are likely to show lower coherency, as they may relate to completely different aspects of the query. For a less ambiguous query, such as \textit{"Tell me about folk remedies for a sore throat"}, we expect the retrieved items to show higher coherency, better reflecting a single intent. To measure coherency, we consider retrieved items as fragments of a larger whole, e.g., passages of a document that are contextually related.  We build a \emph{coherency network} based on the coherency of retrieved items. The graph connectivity measures in the coherency network are considered  as indicators of query ambiguity. The nodes of our coherency graph  are the retrieved items (passages), with vertex $A$ connected to vertex $B$ if passage $B$ is predicted to be likely successor to passage $A$ according to a given language model:

\begin{definition}
\label{ego_def}
Given a set of retrieved items $d_i \in D$ ,  a coherency-network $G(D)$ is denoted as  $G(D)= (\mathbb{V}, \mathbb{E})$, a directed graph, where $
    \mathbb{V}= \{ d_i \in D \}$, and $\mathbb{E} = \{ d_{i} \to  d_{j} : \forall d_i , d_j \in	\mathbb{V} | P_{s}(d_i,d_j)=1\}$.
\end{definition}
The function $P_{s}(d_i,d_j)=1$ when $d_j$ is predicted to be a likely successor to $d_i$; $P_{s}(d_i,d_j)=0$, otherwise.

Figure~\ref{fig:graphs} plots the coherency network for the top-10 retrieved items for four~example queries from the ClariQ dataset. The two examples on the left represent more ambiguous queries, and the two examples on the right represent less ambiguous queries. As shown in this figure, the structure of the coherency network is related to query ambiguity level. The coherency networks for more ambiguous queries are sparser and more scattered, whereas  for less ambiguous queries they are more condensed and connected. Thus, we measure some of the coherency-network properties that would potentially be aligned with this observation.
We measure  the \textbf{Node Connectivity (NC)} of the coherency network as the minimum number of nodes that must be removed to disconnect G or render it trivial\cite{white2001fast}.
The intuition is that if $G$ is prune to get disconnected easily, there are nodes with low interactions with other graphs, which indicates that retrieved items are referring to different topics that do not relate with each other.
Further, we measure \textbf{Average Node Connectivity (ANC})  which is the average of \textit{local node connectivity} over all pairs of nodes of $G$ on the coherency network, as an indicator of coherency-network density and subsequently the clarity level of the query \cite{beineke2002average}. The Local Node Connectivity  for two non-adjacent nodes is defined as the minimum number of nodes that must be removed to disconnect the graph~\cite{brandes2005network}. As shown in Figure~\ref{fig:graphs}, the nodes in the coherency-network of an ambiguous query are more susceptible to becoming disconnected as nodes are removed because of having lower interactions between the vertices.
\section{Experiments}

\subsection{Datasets}
We ran our experiments on the ClariQ and AmbigNQ datasets. In the following,  we briefly explain the details of each dataset. For further details, please refer to the original papers.

\textbf{ClariQ\footnote{\url{https://github.com/aliannejadi/ClariQ}}:}
ClariQ is a dataset dedicated to the problem of
asking clarifying questions in open-domain
dialogue systems. It is based on TREC WEB track 2009-2014 data\footnote{\url{trec.nist.gov}} and contains almost 300 search topics. It aims to support research into both``when'' to ask clarifying question and ``what'' clarifying questions to ask. To develop the dataset, human experts were asked to convert the short keyword-based queries into conversational requests. Then two annotators were assigned to rate each query based on the level of clarification required from 1 (the most clear queries) to 4 (the most ambiguous queries). In case of disagreement, a third annotator made the final assessment. 
Since in this work, we want to predict whether we should ask a clarifying question or not,  we interpret the top two levels of query clarity (level 1 and 2) as questions that do not need clarification and the lower two levels (level 3 and 4) as those that require clarification. We train supervised baselines on ClariQ train set, tune them on its development set, and report the performance of our methods, as well as the baselines, on the ClariQ test set.

\textbf{AmbigNQ\footnote{\url{https://nlp.cs.washington.edu/AmbigNQ/}}:}
\citet{min2020ambigqa} introduced AmbigQA, a new task which necessities identifying all possible answers to
an open-domain question, along with disambiguated questions to distinguish them. In addition, they construct AmbigNQ, a dataset with
over 14K high-quality human annotated questions from the NQ-OPEN dataset \cite{kwiatkowski2019natural}
containing diverse types of ambiguity. We use this dataset to examine the generalizability of different query clarity methods. Questions in AmbigNQ differ from questions in ClariQ by requiring short precise answers, often named entities. Unlike ClariQ, AmbigNQ's definition of ambiguity includes different expressions of the same named entity or alternative answers that do not reflect differ interpretations of the question. For example, the question \textit{``Who plays the white queen in alice through the looking glass?''} is considered to be ambiguous since it has been annotated with two answers \textit{`Amelia Crouch (young White Queen)'} and \textit{`Anne Hathaway (adult White Queen)}. If this question appeared in ClariQ, the two answers could be concatenated with each other and the question might not be considered  ambiguous..  Requests in ClariQ are more general and cannot be necessarily  answered  by a short  phrase or an entity.  Moreover, the ClariQ dataset was specifically collected for the purpose of research into clarifying questions. The questions in AmbigNQ are accompanied by different numbers of question-answer pairs. We take the number of possible question-answer pairs as an indicator of the level of question ambiguity. The greater the number of question-answer pairs, the more ambiguous we take the question to be. 

\begin{figure}[t!]
\centering
\includegraphics[width=0.35\textwidth,clip , trim=0cm 0cm 1cm 1cm,scale=0.2 ]{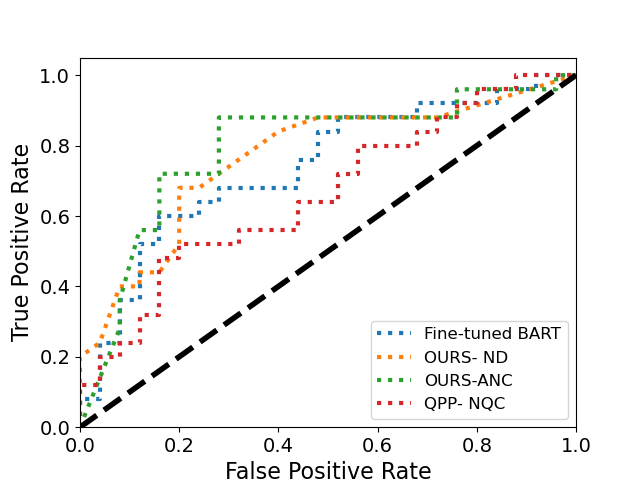}
\caption{Comparing ROC curve for best of QPP methods (NQC), Best of fine-tuned PLM models ( BART) and best of our two proposed methods i.e., BERT-NC and MiniLM-ANC}
\label{fig:roc}
\end{figure}

\subsection{Experimental Setup}
In order to build a coherency-network as described in section \ref{proposed} to determine whether we need to ask clarifying questions, we ought to retrieve passages for a given query. The retrieved items are considered as nodes of the coherency network. We use  the MS MARCO V2 corpus\footnote{\url{https://microsoft.github.io/msmarco/TREC-Deep-Learning-2021}} which is a comprehensive corpus  including more than 138 million passages as the corpus of our retrieval. Further, we employ the well-established BM25 retrieval approach to retrieve passages from the corpus. We selected MS MARCO V2 as the corpus since it has high number of passages and more likely to include relevant answers to each query and BM25 as retrieval methodology since it is well-known approach. Thus, we note that our proposed methodology is not dependent on the corpus nor retrieval methodology. Further, we implement our proposed methods using the NetworkX\footnote{\url{https://networkx.org/}} package to build the coherency network. To build the coherency network, we set $P_{s}(d_i,d_j)=1$ if and only if a pre-trained language model predicts that $d_j$ is next sentence of $d_i$.
As pre-trained models, we used BERT base \cite{devlin2018bert} and also the distilled model MiniLM \cite{DBLP:journals/corr/abs-2002-10957}, which is lighter than BERT while keeping the same level of performance.
Both of these pre-trained models were initially trained on the Masked Language Modeling (MLM) task as well as the \textbf{Next Sentence Prediction (NSP)} task. While MLM teaches the pre-trained language model to understand the relationship between terms, the NSP task goal is to help the model to understand longer-term dependencies and interpret beyond the words to jointly pre-train text-pair representations\cite{devlin2018bert}. Thus, we defined $P_{s}(d_i,d_j)$ on NSP task. Every pair of passages are fed into the model followed by a [SEP] token.  Then, we encode the passages and obtain two probabilities for "IsNextSentence" and "NotNextSentence" class that indicates whether the $d_j$ is predicted to be the successor of  $d_i$. We place an edge from $d_i$ to $d_j$ if the model predicts that $d_j$ is a likely successor of $d_i$. Further, we measure the connectivity metrics explained in section 3 on the coherency-network, we built the coherency network with the top-$k$ retrieved documents were $K\in\{10,20,30,..,100\}$. Among which, using Top-20 documents for building the coherency-network showed the highest performance  on the validation set (dev set) of ClariQ. Thus, we report the performance on top-20 retrieved documents on the test set.

\begin{table}[]
\begin{tabular}{ll||ll||ll}
\multicolumn{2}{c||}{Fine-tuned PLM} & \multicolumn{2}{l||}{QPP Methods} & \multicolumn{2}{l}{Our Proposed methods} \\ \hline
BERT & 0.724 & WIG & 0.552 & \textbf{MiniLm-ANC} & \textbf{0.792} \\
BART &  0.739 & NQC & 0.690 &  BERT-ANC & 0.732 \\
RoBERTa & 0.662 & SMV & 0.680  & MiniLM-NC & 0.757\\
 &  & $n(\sigma_\%)$ & 0.643 &   BERT-NC &  0.767  \\ 

\end{tabular}
\caption{Performance of our QPP methods vs.\ baselines in terms of AUC-ROC on ClariQ test set. }
\label{ATable}
\end{table}

\subsection{Baselines}
We compare the performance of our methods with the SOTA baselines from the original ClariQ paper \cite{aliannejadi2021building}. These baselines fine-tune pre-trained language models based on the (limited) training data in ClariQ to predict the level of query clarity.  We  fine-tuned  RoBERTa \cite{liu2019roberta}, BART \cite{chipman2010bart} and BERT \cite{devlin2018bert} as suggested by \cite{aliannejadi2021building} and tuned the hyper-parameters on the ClariQ dev set by selecting number epochs from \{1,3,5\} and learning rate from $\{5/6/7 e-5/6/7\}$ on ClariQ dev set. By carefully tuning learning rates and number of epochs, we were able to outperform the results reported in the original paper (Table~\ref{ATable}).
Query Performance prediction methods have  shown to be highly correlated with retrieval performance on different well-known benchmarks. Some of these methods are based on the intuition that if a query is ambiguous, it is more likely to have lower performance compared to a more specific one. Thus, 
we considered the SOTA unsupervised QPP methods such as WIG\cite{zhou2007query}, NQC\cite{shtok2012predicting}, SMV\cite{tao2014query} and $n(\sigma_\%)$\cite{cummins2011improved} that have shown relatively consistent performance on different well-known research benchmarks  such as Robust04, TREC Web Collections, and MS MARCO~\cite{carmel2010estimating,hashemi2019performance,arabzadeh2021bert}.
We tuned the hyper-parameters of these QPP methods the on ClariQ dev set,  and report the best-performing models on the test set. 

\subsection{Results}
Since determining "when" to ask clarifying question is a binary classification task, we evaluate our proposed approaches and compared it to baselines in Table~\ref{ATable} in terms of the area under the ROC curve (AOC). As shown in this table, our proposed methods outperform all the QPP methods. We also compared the best of each category, i.e., best of fine-tuned PLM ( BART), best of QPP methods (NQC) and best of our proposed methods from each category, i.e., MiniLM-ANC and BERT-NC. We plot the ROC curves in Figure 3. We note that both MiniLM-ANC and BERT-NC outperform all the other baselines with statistical significance ($p <0 .05$). Our methods outperformed the best when using the top-20  retrieved items. Thus, we report all the other performance measures with the top-20 items. 

\subsection{Generalizability Analysis}

\begin{figure}[]
\centering
\includegraphics[width=0.37\textwidth ]{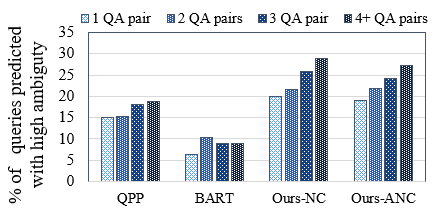}
\caption{Percentage of queries predicted as ambiguous based on each method on 4 different classes of ambiguity in AmbigNQ. }
\label{fig:ambign}
\end{figure}

To determine 
our query clarity prediction methods generalizability
, we evaluate the best performing methods from the ClariQ dataset on the AmbigNQ dataset. Since AmbigNQ was not initially designed for the clarifying question task, we adopt the number of question-answer pairs in AmbigNQ dataset as an indicator of the clarification need level. As reported in the original paper~\cite{min2020ambigqa}, we split the queries in the AmbigNQ dev set into  4 groups, with 1,2,3, and 4+ question answer-pairs.  We expect to see that query clarity measures tend to ask question on level 1 less than level 2, and consequently on level 2 less than level 3, and so on. We ran experiments on the best of the baselines i.e., fine-tuned BART vs. our NQC from QPP methods and  MiniLM-ANC and BERT-NC from our proposed methods and plotted the predicted percentage on ambigNQ dev set. As shown in Figure 4, we expected to see gradually increasing ambiguity across the four classes. Our method fully satisfies this expectation. The other unsupervised baseline, i.e., best of QPP methods also matches this expectation. However, ambiguity does not increase for the supervised PLM-based approaches, which are  trained on the ClariQ training set. This observation, indicates that while supervised baselines showed promising performance, they suffer from lack of generalizability and are biased toward their training data set. 

\section{Conclusion}
In this paper, we demonstrated an unsupervised method to determine when clarifying questions should be asked by conversational systems, based on the coherency of the items retrieved by an initial query. i.e., the more retrieved items are coherent, the less ambiguous the query is and the retrieved items are pointing to the same concept.
We build a directed graph based on retrieved items where the nodes are retrieved items and edges indicate when items are predicted to be likely successors. We showed that  coherency-network connectivity measures are indicators of query clarity and retrieved items coherency and can be used to determine if we need to ask clarifying questions or not.
We validated our methods through experiments on the ClariQ clarifying question dataset and the AmbigNQ question ambiguity dataset, showing that our unsupervised method performed as well or better than supervised methods, as well as well-established query performance prediction baselines. 
We demonstrate that our method outperforms both supervised and unsupervised methods on the ClariQ dataset. 
Due to limited amount of training data available for this task, the generalizability of the approach is important and any bias toward training set should be avoided. Our unsupervised approach demonstrates great generalizability when applied to the AmbigNQ dataset,  compared to SOTA supervised baselines which were trained on ClariQ dataset. 

\bibliographystyle{ACM-Reference-Format}
\balance
\bibliography{acmart} 

\end{document}